\documentclass[twocolumn]{bmcart}

\usepackage{amsthm}
\usepackage{amsmath}
\usepackage[utf8]{inputenc} 
\usepackage{bm}
\usepackage{booktabs}
\usepackage{multirow}
\usepackage{graphicx}
\usepackage{amsfonts,amssymb}
\usepackage{hyperref}
\usepackage[acronym]{glossaries}

\makeglossaries

\startlocaldefs
\endlocaldefs

\begin{document}
\begin{frontmatter}

\begin{fmbox}
\dochead{Research}


\title{Adversarial Joint Training with Self-Attention Mechanism for Robust End-to-End Speech Recognition}


\author[
  addressref={aff1},                   
  corref={aff1},                       
  email={lujun.li@tum.de}   
]{\inits{L}\fnm{Lujun} \snm{Li}}
\author[
  addressref={aff1},
  email={yikai.kang@tum.de} 
]{\inits{Y}\fnm{Yikai} \snm{Kang}} \textsuperscript{$\dagger$}
\author[
  addressref={aff1},
  email={yuchen.shi@tum.de}
]{\inits{Y}\fnm{Yuchen} \snm{Shi}} \textsuperscript{$\dagger$}
\author[
  addressref={aff1},
  email={ludwig.kuerzinger@tum.de}
]{\inits{L}\fnm{Ludwig} \snm{Kürzinger}}
\author[
  addressref={aff1},
  email={tobias.watzel@tum.de}
]{\inits{T}\fnm{Tobias} \snm{Watzel}}
\author[
  addressref={aff1},
]{\inits{G}\fnm{Gerhard} \snm{Rigoll}}


\address[id=aff1]{
  \orgdiv{Department of Electrical and Computer Engineering},             
  \orgname{Technical University of Munich},          
  \city{Munich},                              
  \cny{Germany. \\ \textsuperscript{$\dagger$} Equal contribution, alphabetical order}                                    
}





\begin{abstractbox}

\begin{abstract} 
Lately, the self-attention mechanism has marked a new milestone in the field of automatic speech recognition (ASR). Nevertheless, its performance is susceptible to environmental intrusions as the system predicts the next output symbol depending on the full input sequence and the previous predictions. A popular solution for this problem is adding an independent speech enhancement module as the front-end. Nonetheless, due to being trained separately from the ASR module, the independent enhancement front-end falls into the sub-optimum easily. Besides, the handcrafted loss function of the enhancement module tends to introduce unseen distortions, which even degrade the ASR performance. Inspired by the extensive applications of the generative adversarial networks (GANs) in speech enhancement and ASR tasks, we propose an adversarial joint training framework with the self-attention mechanism to boost the noise robustness of the ASR system. Generally, it consists of a self-attention speech enhancement GAN and a self-attention end-to-end ASR model. There are two highlights which are worth noting in this proposed framework. One is that it benefits from the advancement of both self-attention mechanism and GANs; while the other is that the discriminator of GAN plays the role of the global discriminant network in the stage of the adversarial joint training, which guides the enhancement front-end to capture more compatible structures for the subsequent ASR module and thereby offsets the limitation of the separate training and handcrafted loss functions. With the adversarial joint optimization, the proposed framework is expected to learn more robust representations suitable for the ASR task. We execute systematic experiments on the corpus AISHELL-1, and the experimental results show that on the artificial noisy test set, the proposed framework achieves the relative improvements of 66\% compared to the ASR model trained by clean data solely, 35.1\% compared to the speech enhancement \& ASR scheme without joint training, and 5.3\% compared to multi-condition training. 
\end{abstract}




\end{abstractbox}
\end{fmbox}

\end{frontmatter}


\section{Introduction}
In recent years, attention-based end-to-end neural networks, which subsume the acoustic and language models into a single neural network, trigger the revolution in the field of automatic speech recognition (ASR) \cite{chan2015listen,chorowski2015attention} and are challenging the dominance of Hidden Markov Model-based hybrid systems \cite{hinton2012deep}. Furthermore, the self-attention mechanism has made another breakthrough in the innovation of the attention architecture, which considers the whole sequence at once to model feature interactions that are arbitrarily distant in time, leading to faster convergence and state-of-the-art results in ASR \cite{chiu2018state, povey2018time,tian2019self,salazar2019self,han2019multiself,han2019stateself,pham2019veryself,yeh2019transformerself,luo2020simplifiedself}. The self-attention system predicts the next output symbol conditioned on the full sequence of the previous predictions. Once a mistake occurs in one estimation step due to noise interference, all the subsequent steps will be disturbed. As speech signals are inevitably interfered by various background noises in the realistic environment, it is crucial to improve the robustness of the self-attention mechanism for practical application.

The mainstream solution to the noise robustness problem is adding an independent speech enhancement (SE) module as the front-end of ASR. Speech enhancement aims to transform the interfered speech to its original clean version, which is achieved by various approaches, i.e., the statistical method like Wiener filter \cite{lim1978all}, the time-frequency masking \cite{narayanan2013ideal,wang2014training,nie2015two}, the signal approximation \cite{weninger2014discriminatively,erdogan2015phase}, the spectral mapping \cite{xu2014regression,nie2018deep}, etc. No matter what approach the speech enhancement model adopts to achieve the goal, it is trained separately from the ASR model on different loss functions (i.e., mean squared error \cite{ephraim1990minimum}) and being evaluated by different objective criteria (i.e., Mean Opinion Score (MOS) prediction of the intrusiveness of background noise \cite{hu2007evaluation}, Segmental SNR \cite{quackenbush1995objective}). This mismatch between the enhancement training and the final ASR task leads to a sub-optimum easily \cite{seltzer2008bridging}. Moreover, the handcrafted loss functions tend to generate over-smoothed spectra or introduce unseen distortions, which sometimes even degrade the downstream ASR performance \cite{wang2016joint}.  

To obtain the optimum and circumvent introducing unnecessary distortion, the idea of a joint training framework is proposed for robust speech recognition \cite{Wang2015JointTO,wang2016joint,ochiai2017multichannel,bin2019jointly}. The fundamental concept of the joint training is concatenating the speech enhancement front-end and a downstream ASR model to build an entire neural network and jointly adjust the parameters in each module. The goal here is that the enhancement front-end tends to produce enhanced features desired by the ASR component, and the ASR module can guide the enhancement module to a more discriminative direction. In this way, the joint framework is optimized on the final ASR objectives, i.e., word/character error rate (W/CER). 

Generative adversarial networks (GANs) aim at mapping samples $\hat{x}$ from the distribution $\mathcal{\hat{X}}$ to samples $x$ from another distribution $\mathcal{X}$. There are two components within GANs. One is the generator (G), which performs the mapping; and the other is the discriminator (D), which guides the training of the generator. GANs have been applied to various speech signal processing tasks, such as speech enhancement \cite{pascual2017segan, soni2018time}, robust speaker verification \cite{michelsanti2017speaker}, spoken language identification \cite{shen2017spoken}, speech emotion recognition \cite{sahu2018emotion}, data augmentation \cite{hu2018augmentation}, and robust speech recognition \cite{donahue2018exploring}. 

Inspired by the advancement of self-attention mechanism and various applications of GAN in speech-related tasks, we propose an adversarial joint training framework with self-attention mechanism to boost the robustness of the self-attention ASR systems, which consists of a self-attention speech enhancement GAN (SA\_SEGAN) and a self-attention end-to-end ASR model (SA\_ASR), where we experiment with Transformer \cite{dong2018speech} and Conformer \cite{gulati2020conformer}. The discriminant component of SA\_SEGAN is first utilized to distinguish the enhanced features from the original clean features, instructing the enhancement module to output the clean distribution. When it comes to the stage of the joint training, the D component acts as the global training guide, and it will shift the direction for the G component to produce more congruous features for the ASR task. As the global guide, the discriminator is expected to remedy the limitation of the separate training and handcrafted loss functions, alleviate the distortion, and lead the speech enhancement component to the global optimum. Meanwhile, the enhancement module is supposed to capture more underlying structural characteristics. With this global guide, the whole framework is expected to learn more robust representations compatible with the ASR task automatically. 

In summary, the main contributions of this paper are the following:
\begin{itemize}
    \item We propose a self-attention based jointly-trained adversarial framework targeting robust speech recognition. This framework benefits from the advancement of both self-attention mechanism and adversarial training;
    \item We exert the global adversarial training, where the discriminant component does not concentrate on the enhancement front-end exclusively, but also plays the role of the global training guide.
    \item The proposed framework yields remarkable results, which achieve relative improvements of 66\% compared to the ASR model trained by clean data solely, 35.1\% compared to the scheme without joint training, and 5.3\% compared to multi-condition training.
\end{itemize}

\section{Related Work}
GANs have been applied in speech enhancement tasks without attention \cite{pascual2017segan,phan2020improvingsegan,baby2020isegan} and with attention \cite{phan2020self,koizumi2020speechselfmultiple}. These works validate the functionality of GAN in the enhancement task on diverse objective criteria; however, they lack proofs of the effectiveness of their work for the downstream ASR task. 

GANs have also been employed to improve the robustness of the ASR model \cite{donahue2018exploring, sriram2018robust,Wang2018dereverberation,liu2018boosting}. A potential limitation lies in the weak matching and communication between the integrated modules. For instance, speech enhancement and speech recognition are often designed independently, and the enhancement system is tuned according to metrics that are not straightly relative to the final ASR performance.  


To address this concern, joint training is a promising approach. An early attempt was proposed in \cite{droppo2006joint}, where a feature extraction front-end and a Gaussian Mixture Model-Hidden Markov Model back-end are jointly trained on maximum mutual information. Afterwards, other interesting works are published in this field \cite{wang2016joint,Wang2015JointTO,gao2015joint,ravanelli2016batchnormalizedjoint,qian2016neuraljoint}. Nevertheless, an effective integration between the various systems has been difficult for many years, mainly due to the different nature of the technologies involved at different steps. For example, in \cite{wang2016joint,gao2015joint}, the joint training is actually performed as a fine-tuning procedure. To tackle this problem, this paper deploys the discriminant component of GAN as a global guide, leading the enhancement module to match the downstream ASR module. 


\section{Self-attention Based SE-ASR Scheme}
\subsection{Overview}
\begin{figure}[tb]
\centering
\includegraphics[scale=0.33]{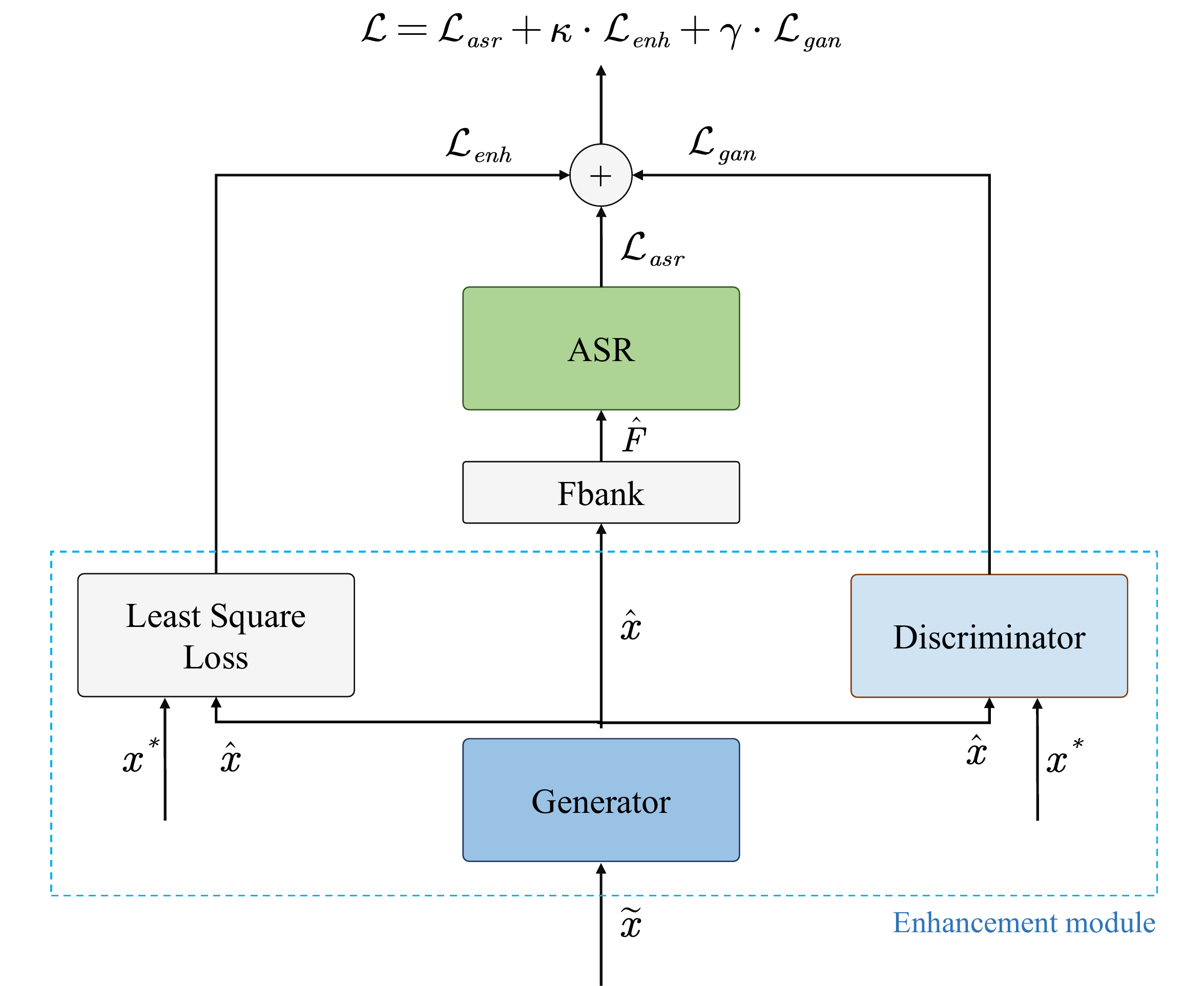}
\caption{Overview of the SE\_ASR joint training framework.}
\label{overview}
\end{figure}

Fig. \ref{overview} illustrates an overview of our proposed joint training framework for robust end-to-end speech recognition pictorially. The system consists of a self-attention enhancement front-end and a self-attention ASR model. Given the raw noisy speech input $\tilde{X}$ and the raw clean input $X^*$, we illustrate the entire procedure of the joint training pipeline in the following forms:
\begin{equation}
\hat{X}=\text{Generator}(\tilde{X}),
\end{equation}
\begin{equation}
\hat{F}=\text{FBank}(\hat{X}),
\end{equation}
\begin{equation}
P(Y|\hat{F})=\text{SA\_ASR}(\hat{F}),
\end{equation}
\begin{equation}
P(D|\hat{X},X^*)=\text{Discriminator}(\hat{X},X^*).
\end{equation}
Here, $\text{Generator}(\cdot)$ acts as a speech enhancement front-end realized by the generator component of SA\_SEGAN \cite{phan2020self}, which transforms the noisy raw input $\tilde{X}$ to the enhanced $\hat{X}$. $\text{FBank}(\cdot)$ is a function for extracting the normalized log FBank features $\hat{F}$ from the enhancement outputs $\hat{X}$. Subsequently, $\text{SA\_ASR}(\cdot)$ is an ASR system based on self-attention layers realized by Transformer \cite{dong2018speech} or Conformer \cite{gulati2020conformer} architecture. $Y$ is the outputs of the whole scheme. $\text{Discriminator}(\cdot)$ is realized by the discriminator component of SA\_SEGAN \cite{phan2020self}, which distinguishes enhanced outputs from clean data.

\subsection{Self-attention Mechanism}
Self-attention \cite{vaswani2017attention} relates the information over different positions of the entire input sequence for computing the attention distribution using scaled dot-product attention:
\begin{equation}
\text{Attention}(\bm{Q}, \bm{K}, \bm{V})=\text{softmax}(\frac{\bm{QK}^T}{\sqrt{d_k}})\bm{V}.
\end{equation}
$\bm{Q}\in \mathbb{R}^{t_q \times d_q}$, $\bm{K}\in \mathbb{R}^{t_k \times d_k}$, and $\bm{V}\in \mathbb{R}^{t_v \times d_v}$ are three inputs of the self-attention layer: queries, keys, and values, where $t_q$, $t_k$, and $t_v$ are the element numbers in different inputs while $d_q$, $d_k$, and $d_v$ denote the corresponding element dimensions. The scalar $\frac{1}{\sqrt{d_k}}$ prevents the softmax function from falling into regions with tiny gradients. One query's output is computed as a weighted sum of the values, where each weight of the value is computed by a designated function of the query with the corresponding key. 
\subsection{Self-attention Speech Enhancement GANs}
\subsubsection{Speech Enhancement GANs (SEGAN)}
Given a dataset $\mathcal{X}=\{(\bm{x^*_1},\bm{\tilde{x}_1)},(\bm{x^*_2},\bm{\tilde{x}_2}),\cdots,(\bm{x^*_N},\bm{\tilde{x}_N})\}$ consisting of $N$ pairs of raw signals: clean speech signal $\bm{x}^*$ and noisy speech signal $\bm{\tilde{x}}$. Speech enhancement aims to find a mapping $f_\theta (\bm{\tilde{x}}):\bm{\tilde{x}}\to \bm{\hat{x}}$ to transform the raw noisy signal $\bm{\tilde{x}}$ to the enhanced signal $\bm{\hat{x}}$. $\theta$ contains the parameters of the enhancement network.

Conforming to GAN's principle \cite{goodfellow2014generative}, the generator G is for learning an effective mapping that can imitate the real data distribution to generate novel samples related to those of the training set. Hence G acts as the enhancement function. In contrast, the discriminator D plays the role of a classifier which distinguishes the real sample, coming from the dataset that G is imitating, from the fake samples, made up by G. D guides $\theta$ towards the distribution of clean speech signals. To sum up, SEGAN designates the generator G for the enhancement mapping, i.e. $\bm{\hat{x}}=G(\bm{\tilde{x}})$, while designates the discriminator D to guide the training of G by classifying $(\bm{x^*},\bm{\tilde{x}})$ as real and $(\bm{\hat{x}},\bm{\tilde{x}})$ as fake. Eventually, G learns to produce enhanced signals $\bm{\hat{x}}$ good enough to fool D such that D classifies $(\bm{\hat{x}},\bm{\tilde{x}})$ as real. 

\begin{figure}[tb]
\centering
\includegraphics[scale=0.215]{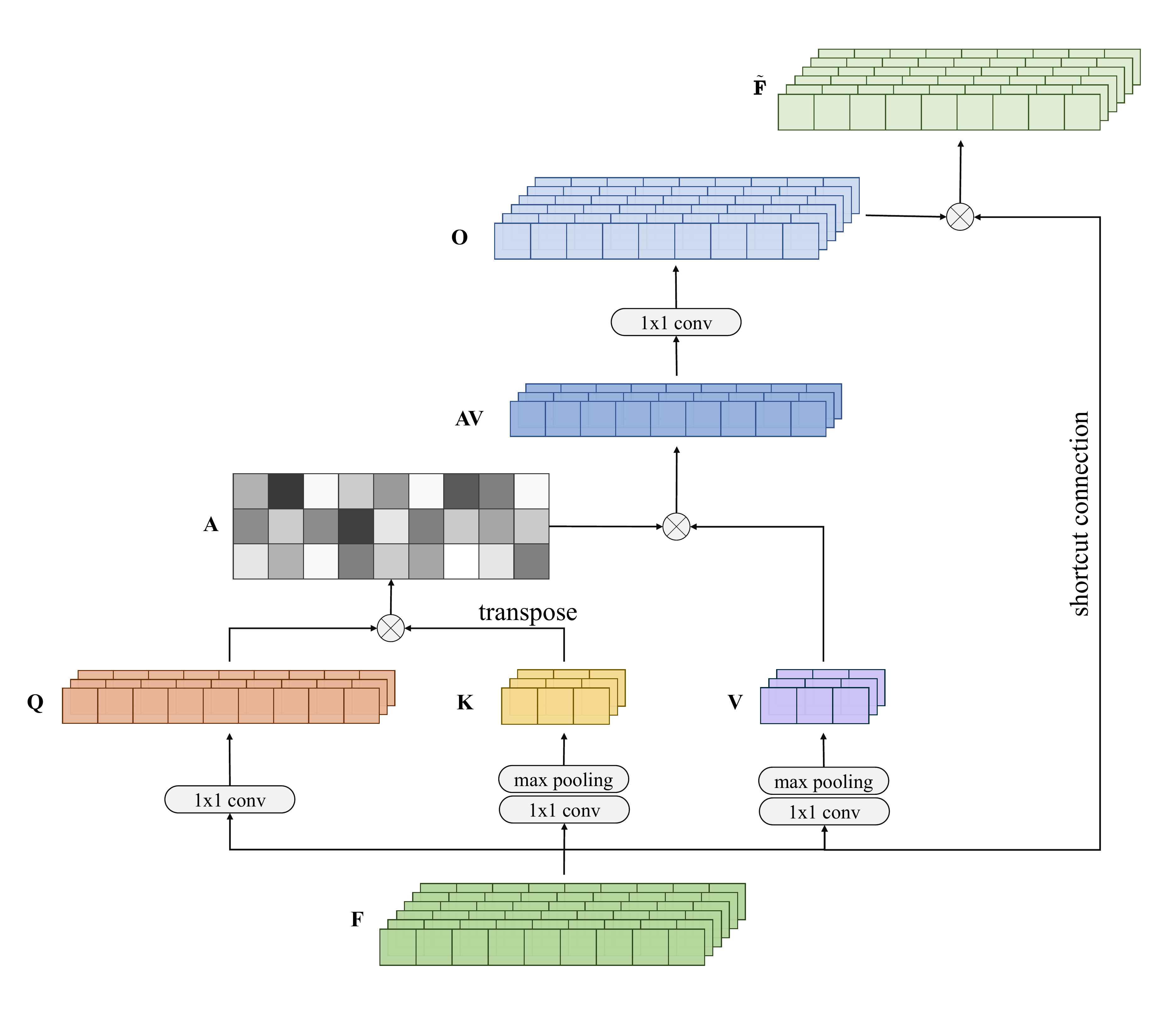}
\caption{Illustration of the application of self-attention mechanism in speech enhancement GANs with $L=9$, $C=6$, $p=3$, and $b=2$. }
\label{sasegan}
\end{figure}

\subsubsection{Self-attention Speech Enhancement GANs (SA\_SEGAN)} \label{SAlayer}
SA\_SEGAN \cite{phan2020self} is SEGAN with the adoption of the self-attention layer adapted from non-local attention \cite{wang2018non,zhang2019self}. Given the feature map $\bm{F}\in \mathbb{R}^{L\times C}$ output by the 1-dim convolutional layer, where $L$ is the time dimension, $C$ is the number of channels, the query matrix $\bm{Q}$, the key matrix $\bm{K}$, and the value matrix $\bm{V}$ are obtained via transformations:
\begin{equation}
\bm{Q}=\bm{FW}^Q,\bm{K}=\bm{FW}^K,\bm{V}=\bm{FW}^V,
\end{equation}
where $\bm{W}^Q$, $\bm{W}^K$, and $\bm{W}^V$ denote the weight matrices of the convolutional layer. Furthermore, Phan et al. \cite{phan2020self} introduce two factors, $b$ and $p$, for memory efficiency. $b$ reduces the channel dimension, while $p$ reduces the number of keys and values by a max pooling layer with filter width and stride size of $p$. Therefore, the dimension of the matrices are $\bm{Q}\in \mathbb{R}^{L\times \frac{C}{b}}$, $\bm{K}\in \mathbb{R}^{\frac{L}{p}\times \frac{C}{b}}$, and $\bm{V}\in \mathbb{R}^{\frac{L}{p}\times \frac{C}{b}}$. The attention map $\bm{A}$ and the attentive output $\bm{O}$ are then computed as 
\begin{equation}
\bm{A}=softmax(\bm{QK}^T),\quad\bm{A} \in \mathbb{R}^{L\times \frac{L}{p}},
\end{equation}
\begin{equation}
\bm{O}=(\bm{AV})\bm{W}^O,\quad \bm{W}^O\in \mathbb{R}^{\frac{C}{b}\times C}.
\end{equation}
Each element $a_{ij}\in \bm{A}$ indicates the extent to which the model attends to the $j$th column $\bm{v}_j$ of $\bm{V}$ when producing the $i$th output $\bm{o}_i$ of $\bm{O}$. With the weight matrix $\bm{W}^O$ realized by a 1 $\times$ 1 convolution layer of $C$ filters, the shape of $\bm{O}$ is restored to the original shape $L \times C$.

In the end, SA\_SEGAN contains a shortcut connection to facilitate information propagation, and a learnable parameter $\beta$ is employed to balance the weight between the output $\bm{O}$ and the input feature map $\bm{F}$ as
\begin{equation}
\bm{F'}=\beta \bm{O}+\bm{F}.
\end{equation}
We illustrate the diagram of a simplified self-attention layer with $L=9$, $C=6$, $p=3$, and $b=2$ in Fig. \ref{sasegan}.


\begin{figure}[tb]
\centering
\includegraphics[scale=0.4]{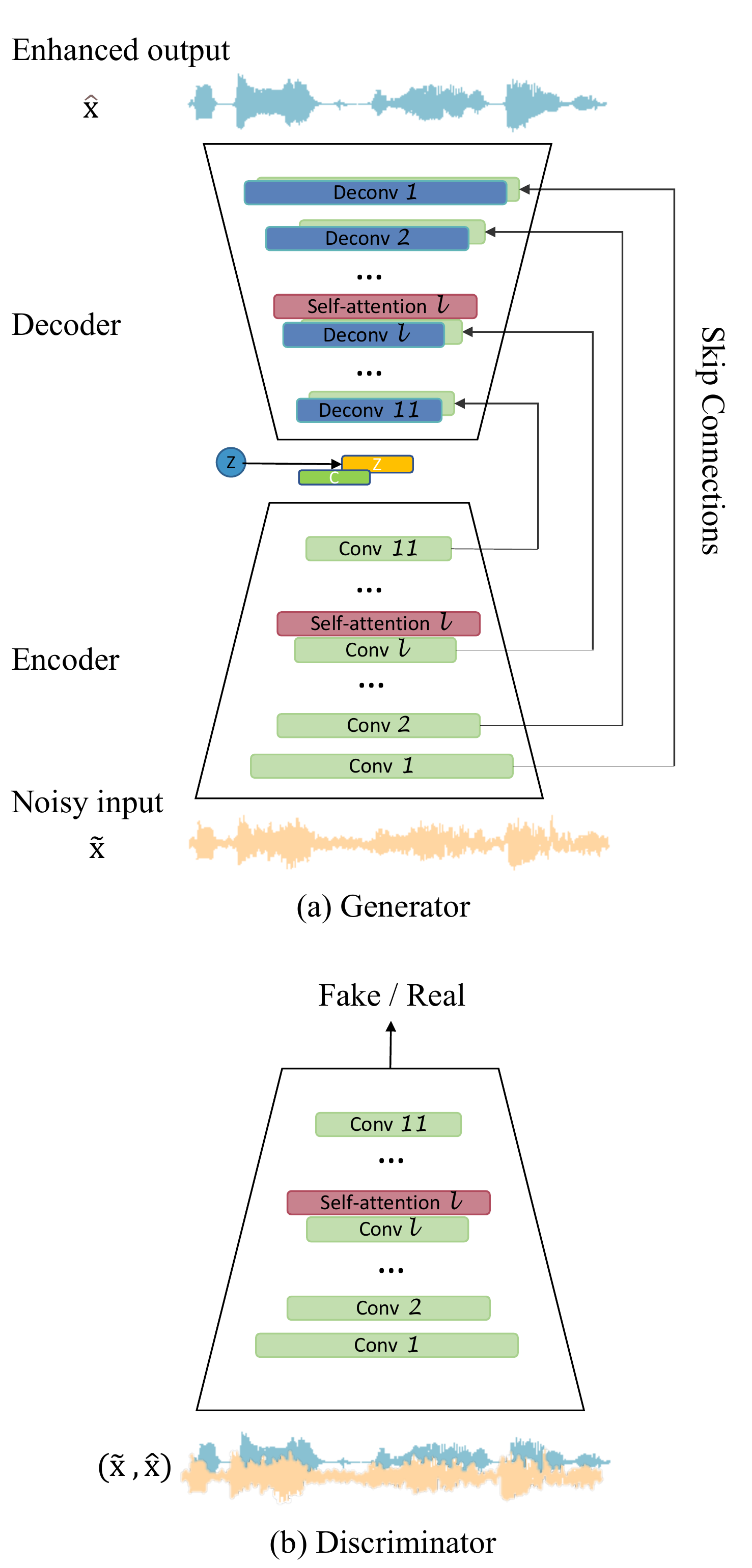}
\caption{Illustration of the SA\_SEGAN architecture. (a) the generator component. (b) the discriminator component. \cite{phan2020self}}
\label{sasegan2}
\end{figure}

\subsubsection{Network Architecture}
The architectures of the generator $G$ and the discriminator $D$ are depicted in Fig. \ref{sasegan2}\,(a) and (b). The G component makes use of an encoder-decoder architecture with fully-convolutional layers \cite{radford2015unsupervised}. The generator's encoder comprises 11 1-dim stridden convolutional layers with a common filter width of 31 and a stride length of 2, followed by parametric rectified linear units (PReLUs) \cite{he2015delving}. The encoder receives a one-second segment of the raw signal sampled at 16 kHz, approximately 16384 samples as the input. To compensate for the smaller and smaller convolutional output, the number of filters increases along the encoder's depth $\{ 16, 32, 32, 64, 64, 128, 128, 256, 256, 512, 1024 \}$ , resulting in output size of the feature map $\{ 8192\times 16, 4096\times 32, 2048\times 32, 1024\times 64, 512\times 64, 256\times 128, 128\times 128, 64\times 256, 32\times 256, 16\times 512, 8\times 1024 \}$. At the 11th layer of the encoder, the encoding vector $\bm{c}\in \mathbb{R}^{8\times 1024}$ is stacked with the noise sample $\bm{z}\in \mathbb{R}^{8\times 1024}$, sampled from the distribution $\mathcal{N}(0, I)$, and presented to the decoder. 

The decoder component mirrors the encoder architecture with the same number of filters and the filter width to reverse the encoding process through deconvolutions. The same as the encoder, each deconvolutional layer is again followed by a PReLUs. The skip connections are deployed to connect the encoding layer with its corresponding decoding layer to allow the information flow between the encoding stage and the decoding stage.

The discriminator is constructed of a similar architecture to the encoder component of the generator. However, it receives the two-channel input and utilize virtual batch-norm \cite{salimans2016improved} before LeakyReLU \cite{maas2013rectifier} activation with $\alpha$ = 0.3. Moreover, the D network is topped up with a $1 \times 1$ convolutional layer to reduce the dimension of the output of the last convolutional layer from $8 \times 1024$ to 8 for the subsequent classification task with the softmax layer.

The self-attention layer illustrated in section \ref{SAlayer} couples with the (de)convolutional layer of both the generator and the discriminator. Fig. \ref{sasegan2} (a) and (b) demonstrate an example of the self-attention layer coupling with the $l$th (de)convolutional layer. As we can see, if we add the self-attention layer to the $l$th convolutional layer of the encoder, the mirror $l$th deconvolutional layer of the decoder and the $l$th layer in the discriminator also couples a self-attention layer. Theoretically, the self-attention layer can be placed in any number, even all, of the (de)convolutional layers.



\subsection{FBank Extraction Network}
We extract the normalized log FBank features $\bm{\hat{f}}$ as the input of the subsequent ASR model, which is computed from the enhanced signals $\bm{\hat{x}}$:
\begin{equation} \label{fbank}
\bm{\hat{f}}=\text{FBank}(\bm{\hat{x}})=\text{Norm}(\text{log}(\text{Mel}(\text{STFT}(\bm{\hat{x}})))),
\end{equation}
where STFT($\cdot$) is the operation of short-time Fourier transform (STFT), Mel($\cdot$) is the operation of Mel matrix multiplication, and Norm($\cdot$) is for normalizing the mean and variance to 0 and 1, separately. Consequently, the FBank feature extraction layer is differentiable.
\subsection{Transformer} \label{transformersection}
\subsubsection{Multi-head Attention Mechanism} \label{multiheadattention}
Multi-head attention mechanism \cite{vaswani2017attention}, as the terminology implies, contains more than one self-attention module. As the core module of the Transformer \cite{dong2018speech}, it leverages different attending representations jointly. Before performing each attention, three linear projections transform the queries, keys, and values to more discriminated representations, respectively. Afterwards, each dot-product attention is calculated independently, and their outputs are concatenated and fed into another linear projection to obtain the final $d_{model}$-dimensional outputs:
\begin{multline}
    \text{MultiHead}(\bm{Q}, \bm{K}, \bm{V}) \\ = \text{Concat}(head_1, head_2,\cdots,head_h)\bm{W}^{OUT},
\end{multline}
where
\begin{equation}
head_i=\text{Attention}(\bm{QW}^Q_i, \bm{KW}^K_i, \bm{VW}^V_i).    
\end{equation}
$h$ refers to the head numbers, and $\bm{Q}$, $\bm{K}$, $\bm{V}$ have the same dimensions of $d_{model}$. Four projection matrices $\bm{W}^Q_i\in \mathbb{R}^{d_{model}\times d_q}$, $\bm{W}^K_i\in \mathbb{R}^{d_{model}\times d_k}$, $\bm{W}^V_i\in \mathbb{R}^{d_{model}\times d_v}$, and $\bm{W}^{OUT}\in \mathbb{R}^{hd_v\times d_{model}}$. Additionally, $d_q=d_k=d_v=d_{model}/h$.
\subsubsection{Positional Encoding}\label{positionalencoding}
One obvious limitation of the Transformer model is that the output is invariant to the input order permutation, i.e., the Transformer does not model the order of the input sequence. Vaswani et al. \cite{vaswani2017attention} solve this problem by injecting information about absolute positions into the input sequence via sinusoid positional embeddings:
\begin{equation}
    PE_{(pos,i)}=\left\{\begin{matrix}
sin(pos/10000^{i/d_{model}}))\quad \text{if}\:\,i\:\, \text{is}\:\, \text{even} & \\ cos(pos/10000^{i/d_{model}}))\quad \text{if}\:\,i\:\, \text{is}\:\, \text{odd} & 
\end{matrix}\right.
,
\end{equation}
where $pos$ refers to the position and $i$ is the dimension. The sinusoidal function allows the model to extrapolate from long sequence lengths.
\subsubsection{Feed-forward Network} \label{feedforwardnetwork}
The feed-forward network (FFN) is another core module of the Transformer \cite{dong2018speech}. It is composed of two linear transformations with a ReLU activation in between. The dimensionality of the input and output is $d_{model}$, and the inner layer has the dimensionality $d_{ff}$. Specifically, 
\begin{equation} \label{ffntransformer}
    \text{FFN}(\bm{x})=\text{max}(0, \bm{x}\bm{W}_1+\bm{b}_1)\bm{W}_2+\bm{b}_2,
\end{equation}
where the weights $\bm{W}_1\in \mathbb{R}^{d_{model}\times d_{ff}}$, $\bm{W}_2\in \mathbb{R}^{d_{ff}\times d_{model}}$ and the biases $\bm{b}_1\in \mathbb{R}^{d_{ff}}$, $\bm{b}_2\in \mathbb{R}^{d_{model}}$. The linear transformations are the same across different positions.

\begin{figure}[tb]
\centering
\includegraphics[scale=0.45]{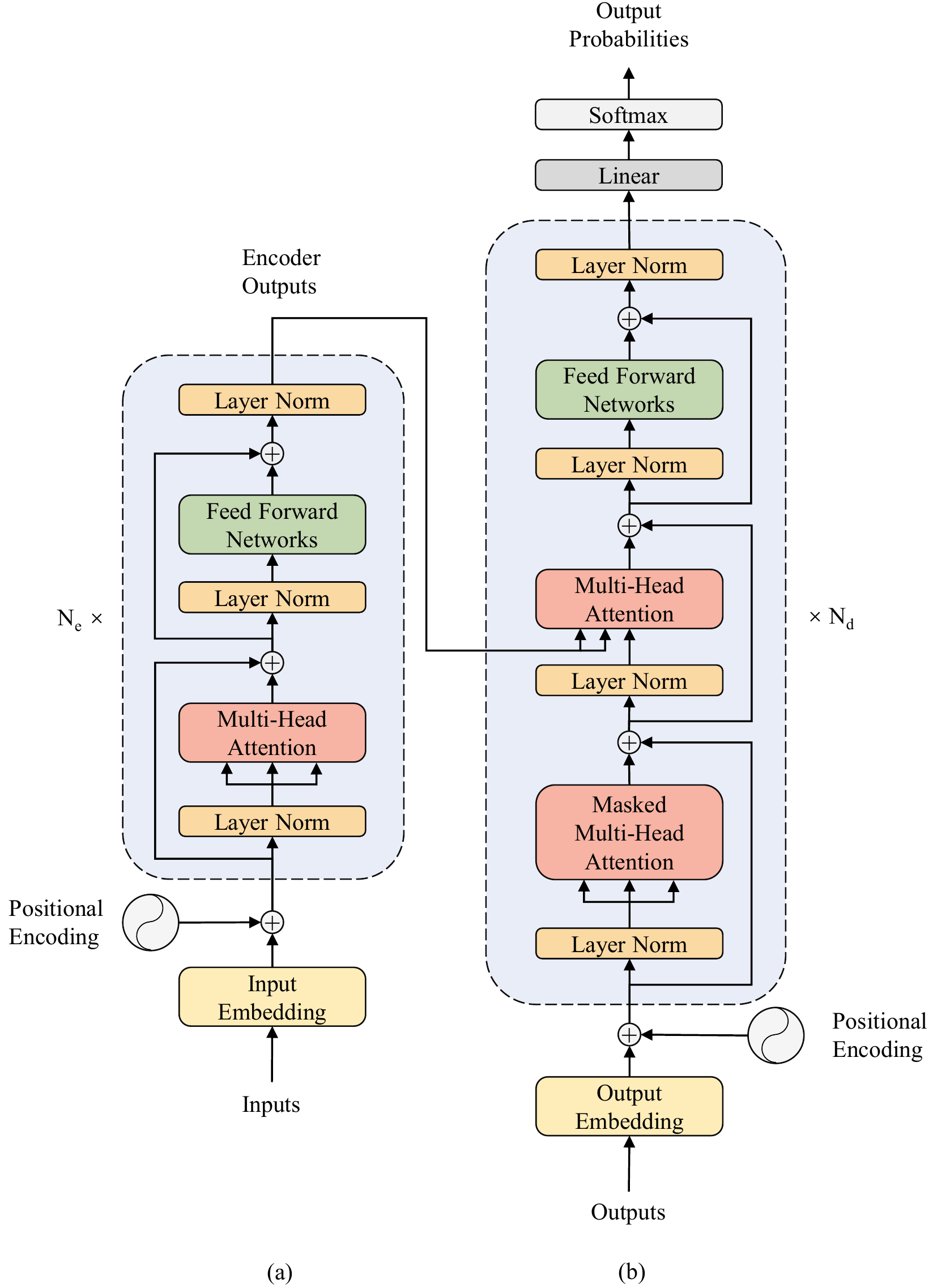}
\caption{Model architecture of the Transformer. (a) Encoder (b) Decoder \cite{dong2018speech}}
\label{transformer}
\end{figure}

\subsubsection{Network Architecture}
The detailed model architecture of the ASR-Transformer is as follows:

The encoder is shown in Fig. \ref{transformer}\,(a). The input-embedding is for extracting expressive representations of dimension $d_{model}$. Thereafter, to enable the model to attend on the auxiliary position information, the $d_{model}$-dim positional encoding (Section \ref{positionalencoding}) is added to the input encoding. Then the sum of encoded outputs is fed into a stack of $N_e$ encoder blocks, each of which has two sub-blocks: one is the multi-head attention (Section \ref{multiheadattention}), receiving queries, keys, and values from the previous block; the other is the feed-forward networks (Section \ref{feedforwardnetwork}). In the meanwhile, layer normalization and residual connection are introduced to each sub-block for effective training. Thus, the pipeline of the sub-block is:
\begin{equation}
    \bm{x}+\text{SubBlock}(\text{Layer\,Norm}(\bm{x})).
\end{equation}

The decoder is shown in Fig. \ref{transformer}\,(b). The output-embedding converts the character sequence to dimension $d_{model}$. Added with the positional encoding, the sum of them is fed into a stack of $N_d$ decoder blocks, which consists of three sub-blocks: The first is a masked multi-head attention, which ensures that the predictions for position $j$ depends only on the known outputs at positions less than $j$. The second is a multi-head attention whose keys and values come from the encoder outputs while queries come from the previous sub-block outputs. The third is also feed-forward networks. Similar to the encoder, layer normalization and residual connection are also employed to each sub-block of the decoder. Eventually, the output probabilities are acquired by a linear projection and a subsequent softmax function.

\begin{figure*}[tb]
\centering
\includegraphics[scale=0.32]{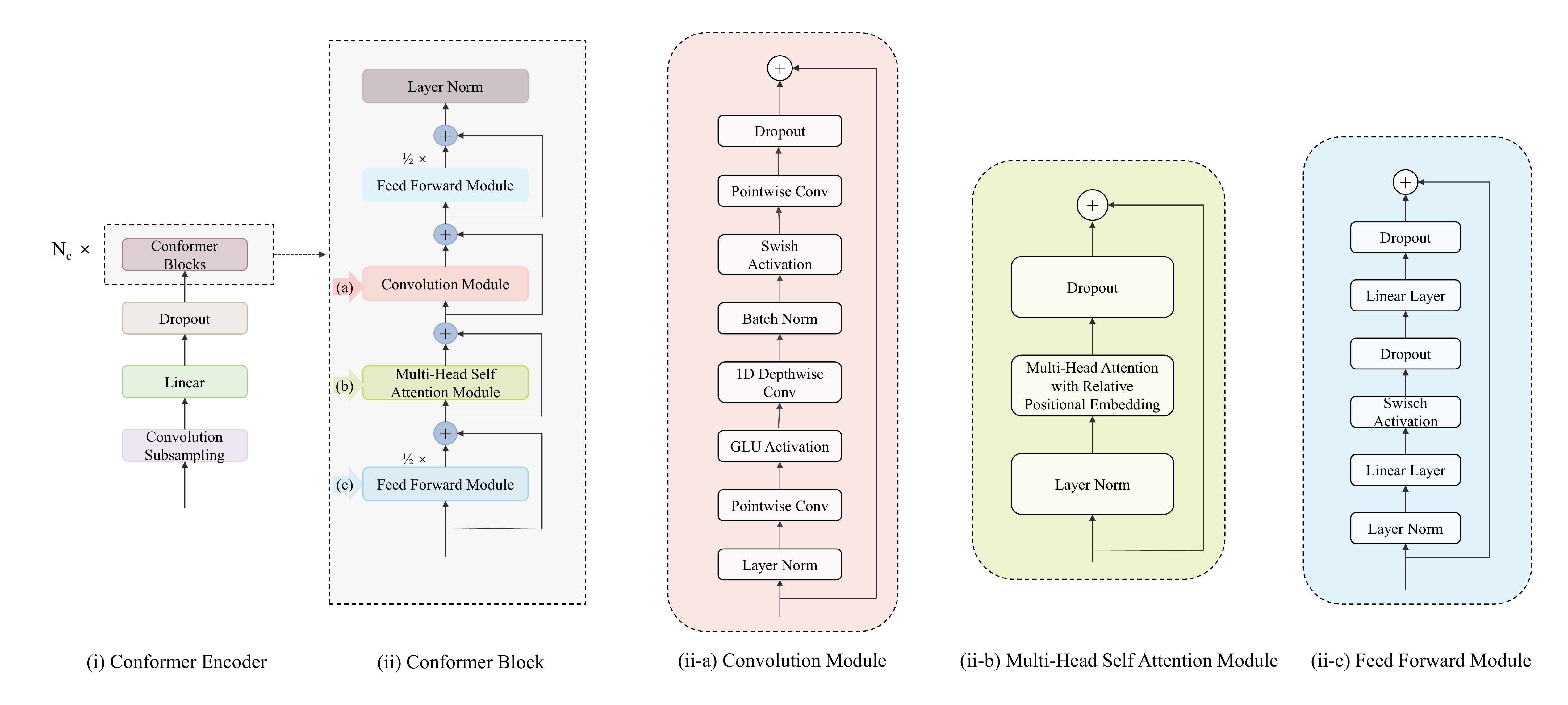}
\caption{Illustration of the Conformer encoder model architecture. (i) Conformer encoder architecture. (ii) Conformer block architecture. (ii-a) Convolution module of the Conformer block. (ii-b) Multi-headed self-attention module of Conformer block. (ii-c) Feed forward module of Conformer block.}
\label{conformer}
\end{figure*}

\subsection{Conformer}
Conformer \cite{gulati2020conformer} is a state-of-the-art ASR encoder architecture. Different from the Transformer block (as described in Section \ref{transformersection}), it is equipped with a convolution layer to increase the local information modeling capability of the Transformer encoder model \cite{vaswani2017attention} and a pair of FFN modules sandwiching the multi-head self-attention module and the integrated convolution module. The Conformer model consists of a Conformer encoder proposed in \cite{gulati2020conformer} and a Transformer decoder \cite{dong2018speech}. The encoder first processes the input with a convolution subsampling layer and then with Conformer blocks, as illustrated in Fig. \ref{conformer} (i). The Conformer block (Fig. \ref{conformer} (ii)) consists of a multi-head self-attention module (MHSA), a convolution module, sandwiched by a pair of macron-feedforward module \cite{lu2019understanding}. The layer normalization is applied before each module and the dropout is followed by a residual connection afterwards (pre-norm) \cite{zeyer2019comparison,wang2019learning}. Mathematically, let $\bm{x}_i$ be the input to the $i$th Conformer block, the output $\bm{y}_i$ of this block is:

\begin{gather}
    \bm{x}'_i=\bm{x}_i+\frac{1}{2}\text{FFN}(\bm{x}_i),  \\
    \bm{x}''_i=\bm{x}'_i+\text{MHSA}(\bm{x}'_i),  \\
    \bm{x}'''_i=\bm{x}''_i+\text{Conv}(\bm{x}''_i), \\
    \bm{y}_i=\text{Layer\,Norm}(\bm{x}'''_i+\frac{1}{2}\text{FFN}(\bm{x}'''_i)).
\end{gather}
$\text{FFN}(\cdot)$, $\text{MHSA}(\cdot)$, $\text{Conv}(\cdot)$, and $\text{Layer\,Norm}(\cdot)$ denote the macron-feedforward module, the multi-head self-attention module, the convolution module, and the layer normalization module, respectively. The multi-head self-attention module is the same as in Section \ref{multiheadattention} and is demonstrated in Fig. \ref{conformer} (ii-b). Section \ref{convolutionmodule} and \ref{ffwmodule} introduce the convolution module and the macron-feedforward module, respectively. 

\subsubsection{Convolution Module} \label{convolutionmodule}
Fig. \ref{conformer}\,(ii-a) demonstrates the details of the convolution module. The convolution module starts with a 1-dim pointwise convolution layer and a gated linear units (GLU) activation \cite{dauphin2017language}. The 1-dim pointwise convolution layer doubles the input channels, and the GLU activation splits the input along the channel dimension and executes an element-wise product. What follows are a 1-dim depthwise convolution layer, a batch normalization layer, a Swish activation, and another 1-dim pointwise convolution layer. As mentioned before, the layer normalization is applied before each module and the dropout is followed by a residual connection afterwards (pre-norm).

\subsubsection{Macron-feedforward Module} \label{ffwmodule}
Unlike the FFN module in Transformer encoder \cite{dong2018speech}, which comprises two linear transformations with a ReLU activation in between (Equ. \ref{ffntransformer}), Conformer encoder \cite{gulati2020conformer} introduces another FFN module and substitutes the ReLU activation with the Swish activation. Furthermore, inspired by Macaron-Net \cite{lu2019understanding}, this pair of FFN modules are following a half-step scheme and sandwiching the MHSA and the convolution modules. The detail of the FFN is illustrated in Fig. \ref{conformer}\,(ii-c).


\section{Adversarial Joint Training}
GANs aim at mapping samples $\bm{\hat{x}}$ from the distribution $\mathcal{\hat{X}}$ to samples $\bm{x}^*$ from another distribution $\mathcal{X^*}$. The generator G is tasked to learn an effective mapping that can imitate the real data distribution to generate novel samples from the manifold defined in the $\mathcal{X}$, by means of an adversarial training exerted by the discriminator D. During back-propagation, D classifies real samples from the fake samples more accurately; in return, G updates its parameters towards the real data manifold, till the mixed Nash equilibria are reached \cite{goodfellow2014generative}. The GAN training process can be formulated as a minimax game between G and D, with the objective
\begin{equation} 
\begin{split}
    \min_{G}\: \max_{D}\: \mathcal{L}(D, G) = & \mathbb{E}_{\bm{x^*}\sim P_{data}(\bm{x^*})}[\text{log}\, D(\bm{x^*})]+ \\
    & \mathbb{E}_{\bm{\hat{x}}\sim P_{\hat{x}}(\bm{\hat{x}})}[\text{log}(1-D(G((\bm{\hat{x}}))))].
\end{split}
\end{equation}

In our proposed robust end-to-end speech recognition scheme, the discriminant network first acts as the local guide for the enhancement module, where D shifts the training of G towards the distribution of clean data; thereafter, it is deployed as the global guide for the whole scheme, where D instructs G to output pertinent enhanced data for the subsequent ASR task. 

We first train the enhancement module, which contains both the generator and the discriminator. To solve the problem of vanishing gradients caused by sigmoid cross-entropy loss for training, the least-squares GAN (LSGAN) with binary coding (1 for real, 0 for fake) is utilized instead of the cross-entropy loss. Consequently, the loss function of the discriminator component changes to
\begin{equation} \label{ganloss}
\begin{split}
    \min_{D}\: \mathcal{L}(D)= & \frac{1}{2}\mathbb{E}_{\bm{x^*},\bm{\tilde{x}}\sim P_{data}(\bm{x^*}, \bm{\tilde{x}})}[D(\bm{x^*},\bm{\tilde{x}})-1]^2+ \\
    & \frac{1}{2}\mathbb{E}_{\bm{z}\sim p_{\bm{z}}(\bm{z}), \bm{\tilde{x}}\sim p_{data}(\bm{\tilde{x})}}[D(G(\bm{z},\bm{\tilde{x}}),\bm{\tilde{x}})]^2,
\end{split}
\end{equation}
where $\bm{z}$ is a latent variable. To minimize the distance between its generations and the clean examples, it is beneficial to add a secondary component to the loss of G. Inspired by the effectiveness of $L_1$ norm in the image manipulation domain \cite{isola2017image,pathak2016context}, we deploy it in G component to gain more fine-grained and realistic results. The magnitude of the $L_1$ norm is controlled by a new hyper-parameter $\lambda$. Hence, the loss function of the generator component becomes 
\begin{equation} \label{enhloss}
\begin{split}
    \min_{G}\: \mathcal{L}(G)= & \frac{1}{2}\mathbb{E}_{\bm{z}\sim p_{\bm{z}}(\bm{z}), \bm{\tilde{x}}\sim p_{data}(\bm{\tilde{x})}}[D(G(\bm{z},\bm{\tilde{x}}),\bm{\tilde{x}})-1]^2 \\
    & +\lambda\left \| G(\bm{z},\bm{\tilde{x}})-\bm{x^*}\right \|_1.
\end{split}
\end{equation}

In the joint training, the enhancement module is initialized from the trained G component, while the global discriminant module is initialized from the trained D component. The training of the ASR component is based on the cross entropy criterion, namely
\begin{equation} \label{asrloss}
    \mathcal{L}_{asr}=-\text{ln}P(Y^*|F)=-\sum_{n}\text{ln}P(y_n^*|F,y_{1:n-1}^*),
\end{equation}
where $Y^*$ is the ground truth of a whole sequence of output labels and $y_{1:n-1}^*$ is the ground truth from output step 1 to $n-1$. In the proposed framework, the parameters of all procedures, enhancement, feature extraction, ASR, and the discriminant network, are updated by stochastic gradient descent calculated by the loss function of the whole scheme. It is composed of three losses: $\mathcal{L}_{asr}$, $\mathcal{L}_{enh}$, and $\mathcal{L}_{gan}$, which correspond to Eqs. \ref{asrloss}, \ref{enhloss}, and \ref{ganloss}, i.e.
\begin{equation} \label{jointtraining}
    \mathcal{L}=\mathcal{L}_{asr}+\kappa \mathcal{L}_{enh}+\gamma \mathcal{L}_{gan},
\end{equation}
where $\kappa$ and $\gamma$ are two hyper-parameters weighting the magnitude of the enhancement loss and adversarial loss. Notably, the scheme targets the recognition performance, and the loss function of the discriminant network adapts the enhancement module implicitly. As a result, the discriminant network guides the enhancement module to serve the subsequent ASR task more properly. Accordingly, the unnecessary speech distortion caused by the enhancement process is alleviated.  

\section{Experimental Setups}
We systematically evaluate the robustness of the adversarial joint training framework, and ablation tests are conducted to validate the effects of (i) the enhancement front-end on the ASR task, (ii) the joint training on the whole scheme, and (iii) the GAN on the joint training. 
\subsection{Corpus}
All experiments are executed on the open source Mandarin speech corpus, AISHELL-1 \cite{bu2017aishell}. This corpus is 178h long, and its utterances contain 11 domains, e.g., smart home, autonomous driving,  industrial production, etc. 400 speakers from different accent areas in China participate in the recording. The corpus is divided into training, development, and test sets. The training dataset contains 120,098 utterances from 340 speakers, the development dataset contains 14,326 utterances from 40 speakers, and the test dataset contains 7,176 utterances from 20 speakers.

For the noisy data, we contaminate clean utterances in AISHELL-1 with 9 sorts of intrusions from the NOISEX-92 dataset \cite{varga1992noisex} artificially as noisy utterances. We create noisy training, development, and test sets in the same manner. Note that besides the ``matched" noisy test set, which is contaminated by the same intrusions as the training dataset, we also corrupt the test set with the rest 5 sorts of intrusions in the NOISEX-92 dataset as ``unmatched" test materials. Table \ref{matchunmatch} exhibits the sorts the intrusions mixed in ``match" and ``unmatch" cases. All utterances are mixed with the intrusions at SNRs randomly sampled between [0dB, 20dB]. 
To sum up, we have two sorts of datasets for training: 
\begin{itemize}
    \item clean: Clean utterances from the training dataset of AISHELL-1.
    \item match: Contaminated clean utterances (training dataset) with ``matched" noises of Table \ref{matchunmatch}.
\end{itemize}
For test datasets, we have: 
\begin{itemize}
    \item clean: clean utterances from the test dataset of AISHELL-1.
    \item match: Contaminated clean utterances (test dataset) with ``matched" noises of Table \ref{matchunmatch}.
    \item unmatch: Contaminated clean utterances (test dataset) with ``unmatched" noises of Table \ref{matchunmatch}.
\end{itemize}

\subsection{Baseline}
For the comparison purpose, we take the work from \cite{bin2019jointly} as the baseline model.

In \cite{bin2019jointly}, the mask-based enhancement network is deployed as the front-end. It estimates a masking function to multiply the frequency-domain feature of the noisy speech to form an estimate of the clean speech. For the ASR task, Liu et al. employ the ESPnet model \cite{watanabe2018espnet}. It consists of an encoder network that maps the input feature sequence into a higher-level representation. Then a location-based attention layer integrates the representation into a context vector with the attention weight vector. In the end, the decoder predicts the next output conditioned on the full sequence of previous predictions. Besides, there is an extra discriminant network, whose loss is weighted in the loss function of the whole scheme to optimize the joint training.  

Importantly, the baseline model does not contain any self-attention layer. Furthermore, the discriminant work in the baseline model is an extra auxiliary module, which does not participate in the enhancement training directly. By contrast, our work benefit from self-attention mechanism and the discriminant module exits innately, which is a component of the enhancement front-end. It acts as the local guide for the enhancement training, leading the enhancement network to output towards the distribution of the clean samples. Simultaneously, it also plays the role of the global guide, instructing the enhancement module and the ASR module better matched. 
\begin{table*}[h!]
\caption{The demonstration of categories of intrusions utilized in ``match" and ``unmatch" cases. }
\label{matchunmatch}
\begin{tabular}{ll}
\toprule
\multicolumn{2}{c}{\textbf{match}}            \\ \midrule
Intrusion & Description \\ \midrule
White Noise & Analog noise generator \\
Factory Floor Noise 1     & Plate-cutting and electrical welding                      \\
Cockpit Noise 1   & Buccaneer jet traveling at 190 knots                      \\
Cockpit Noise 3    & F-16                      \\
Engine Room Noise      &Destroyer                      \\
Military Vehicle Noise     & Leopard 1 vehicle                      \\
Machine Gun Noise         & Gun                     \\
Vehicle Interior Noise     & Volvo 340                      \\
HF Channel Noise          & HF radio channel                      \\ \midrule
\multicolumn{2}{c}{\textbf{unmatch}}          \\ \midrule
Intrusion & Description \\ \midrule
Pink Noise  & Analog noise generator                      \\
Factory Floor Noise 2    & Car production hall                      \\
Cockpit Noise 2    & Buccaneer jet traveling at 450 knots                     \\
Operations Room Background Noise   & Destroyer                      \\
Military Vehicle Noise        & M109                      \\ 
\bottomrule
\end{tabular}
\end{table*}

\subsection{Configurations}
\subsubsection{Baseline}
For the enhancement front-end, the input is the 257-dim logarithmic STFT features, and all input vectors are normalized to have the zero mean and the unit variance. The network is composed of 3-layer long short-term memory (LSTM) with 128 nodes, followed by a linear layer with the sigmoid activation function. The network outputs the masking estimate, whose size is equal to the input size, multiplying by the STFT feature of the noisy speech to estimate the clean speech. 

For the ASR network, the input is the 80-dim normalized log FBank features transformed from the enhanced STFT features. The encoder is composed of 4-layer bidirectional LSTM (BLSTM) with 320 cells, while the decoder is composed of 1-layer unidirectional LSTM with 320 cells. After each BLSTM layer, a linear projection layer with 320 nodes is used to combine the forward and backward LSTM outputs. The location-based attention mechanism comprises 10 centered convolution filters of width 100. Besides, We also adopt a joint connectionist temporal classification (CTC)-attention multitask loss function \cite{kim2017joint} with the CTC loss weight as 0.1.

The discriminant network consists of a 4-layer convolution network, each of which is followed by the ReLU activation function \cite{nair2010rectified}. 

For decoding, we use a beam search algorithm with the beam size 12. CTC rescores the hypotheses with 0.1 weight \cite{kim2017joint}. Besides, an external recurrent neural network (RNN) language model is also adopted with 0.2 weight during decoding.
\subsubsection{The Proposed Joint Training Scheme}
\paragraph{SA\_SEGAN}
The SA\_SEGAN is trained for 86 epochs with RMSprop \cite{tieleman2012lecture} and a learning rate of 0.0002. The batch size is 50. During training, we extract 1-second chunks of raw waveforms ($L=16,384$ samples) with a 50\% overlap. During the test, we slide the window without overlapping through the whole duration of our test utterances and concatenate the outputs at the end of the stream. During both training and test, we employ a high-frequency preemphasis filter with a coefficient of 0.95 to all inputs. For the self-attention layer in SA\_SEGAN, we use $b=8$ and $p=4$ for memory reduction. Phan et al. \cite{phan2020self} suggest that the placement of the self-attention layer does not show a clear difference on the performance, which indicates that applying the self-attention to the higher-level (de)convolutional layer is expected to be as good as to a lower layer. Compromising between the computation time and memory requirement and the performance, we place the self-attention layer in the 10th layer ($l$=10).

\paragraph{FBank extraction network}
The FBank feature extraction network is a linear layer to transform the raw outputs from the upstream SA\_SEGAN to the downstream ASR procedure. We extract 80-dim filterbanks with the window size of 25ms and the window shift of 10ms, extended with the temporal first- and second-order differences. Thereafter, we do the logarithmic calculation and global mean and variance normalization according to Eq. \ref{fbank}.

\paragraph{Transformer}
For training the Transformer, we adopt Adam optimizer \cite{kingma2014adam} with $\beta_1 =0.9$, $\beta_2 = 0.98$, $\epsilon = 10^{-9}$, and vary the learning rate over the course of training according to the formula:
\begin{equation}
\begin{aligned}[b]
    lr=k' \cdot d^{-0.5}_{model} \cdot \text{min}(n^{-0.5}, n \times warmup_n^{-1.5})
\end{aligned}
\end{equation}
where $n$ denotes the step number. $k'$ is a tunable scalar, which is set to be 10 initially and is declined to 1 when the model converges. The learning rate increases linearly during the fist $warmup_n = 25000$ steps, and afterwards, it decreases proportionally to the inverse square root of the step number. We apply the residual dropout to each sub-block before adding the residual information, while the attention dropout is performed on the softmax activations in each attention. Both of these aforementioned dropouts are set to be 0.1. Additionally, we guide the system to be more attentive on closer positions by punishing the attention weights of more distant position-pairs. Similar to the baseline model, we also adopt a joint CTC-attention multi-task loss function \cite{kim2017joint}, with the CTC loss weight as 0.3.  In the decoding, we set the beam size to 12 and length penalty $\alpha$ = 1.0 \cite{wu2016google}. Besides, we also integrate an external RNN language model with 0.3 weight. The training procedure is stopped after 30 epochs.

\paragraph{Conformer}
The model hyper-parameters of the Conformer are: $N_e$=12, $N_d$=6, $H$=4, $d_{k}$=256 and $d_{ff}$=2048. The convolution subsampling layer possesses a 2-layer convolutional neural network (CNN) with 256 channels, stride with 2, and kernel size with 3. The kernel size of the convolution module is 31. We apply dropout in each residual unit of the Conformer with the weight 0.1. The same as the Transformer, we train the network with Adam optimizer \cite{kingma2014adam} with $\beta_1 =0.9$, $\beta_2 = 0.98$, $\epsilon = 10^{-9}$ and a Transformer learning rate schedule \cite{vaswani2017attention} with 10000 warm-up steps. The learning rate is peaked at $0.05/\sqrt{d}$, where $d$ is the model dimension in the Conformer encoder. Note that we do not apply speed perturbation \cite{ko2015audio} or SpecAugment \cite{park2019specaugment} for the data augmentation to exclude extra tricks that could cause performance improvements. The training procedure is stopped after 30 epochs.

\section{Results} \label{results}
We use character error rate (CER) to quantify the system performance in all experiments. We report CER of the AISHELL-1 test set on three conditions: ``clean" refers to the original clean test dataset of the corpus, ``match" denotes the noisy test dataset contaminated by ``matched" sorts of intrusions in Table \ref{matchunmatch}, and ``unmatch" means the noisy test set corrupted by the rest of ``unmatched" sorts of intrusions in Table \ref{matchunmatch}. To validate the efficacy of the enhancement front-end, we also introduce multi-condition training (MCT) for comparison, where we artificially contaminate the training dataset of AISHELL-1 with background noise at a certain SNR. Note that there are three randomizations: (i) The utterance to be corrupted is chosen randomly, and in total 90\% of the training utterances are corrupted; (ii) The background noise is chosen from the ``matched" intrusions from Table \ref{matchunmatch} randomly; (iii) The SNR is sampled randomly between [0dB, 20dB]. Therefore, MCT training data comprise 10\% original clean data and 90\% contaminated data, which is corrupted by one of the ``matched" noises at an SNR in the range of [0dB, 20dB].

Firstly, we train the ASR network with the original clean utterance and multi-condition training strategy. The results are shown in Table \ref{table1}.
\begin{table}[h!]
\caption{CER[\%] results of ASR system trained by clean data and multi-condition training (MCT) without the enhancement.}
\label{table1}
\begin{tabular}{ccccc}
\toprule
\multirow{2}{*}{ASR}         & \multirow{2}{*}{Training} & \multicolumn{3}{c}{CER{[}\%{]}} \\ \cline{3-5} 
                             &                           & clean    & match    & unmatch   \\ \midrule
\multirow{2}{*}{baseline}    & clean                     & 14.0       & 60.0       & 61.6      \\
                             & MCT                       & 14.6     & 20.8     & 27.3      \\ \midrule
\multirow{2}{*}{Transformer} & clean                     & 8.0        & 35.6     & 37.1      \\
                             & MCT                       &  7.8        & 13.1         & 16.2          \\ \midrule
\multirow{2}{*}{Conformer}   & clean                     & 6.5      & 32.6     & 31.7      \\
                             & MCT                       & 6.9      & 12.1     & 13.7      \\ \bottomrule
\end{tabular}
\end{table}
Ranking these three models from the aspect of ASR performance, the first is Conformer, then Transformer, and the last is the baseline model, consistent with observations in \cite{dong2018speech,gulati2020conformer}. However, their performance deteriorates rapidly in the noisy test set, demonstrating the necessity of the robustness investigation. The MCT training considerately improves the system's robustness. Its performance on the ``matched" test set outperforms the clean training by 63.2\% and 62.9\% relative, in cases of Transformer and Conformer model respectively; while on the ``unmatched" dataset, it outperforms the clean training by 56.3\% and 56.8\% relative, in cases of Transformer and Conformer model respectively.

\begin{table}[h!]
\caption{The impacts of the enhancement front-end on ASR systems trained by clean data and multi-condition training (MCT). Results are in CER[\%].}
\label{tableenhance}
\begin{tabular}{ccccc}
\toprule
\multirow{2}{*}{ASR}         & \multirow{2}{*}{Training} & \multicolumn{3}{c}{CER{[}\%{]}} \\ \cline{3-5} 
                             &                           & clean    & match    & unmatch   \\ \midrule
\multirow{2}{*}{baseline}    & clean                     & 13.9       & 25.8       & 53.8      \\
                             & MCT                       & 14.9     & 23.5     & 34.3      \\ \midrule
\multirow{2}{*}{Transformer} & clean                     & 8.1        & 19.1     & 33.7      \\
                             & MCT                       &  7.9        & 14.9         & 20.7          \\ \midrule
\multirow{2}{*}{Conformer}   & clean                     & 6.5      & 17.6     & 28.8      \\
                             & MCT                       & 7.0      & 14.2     & 17.9      \\ \bottomrule
\end{tabular}
\end{table}

\begin{table*}[t!]
\caption{CER[\%] results of the SE\_ASR system retraining with and without noisy features. }
\label{retraining}
    \begin{tabular}{ccccc}
    \toprule
    \multirow{2}{*}{ASR}              & \multirow{2}{*}{Retraining} & \multicolumn{3}{c}{CER{[}\%{]}} \\ \cline{3-5} 
                                      &                             & clean    & match    & unmatch   \\ \midrule
    \multirow{3}{*}{Transformer\_MCT} & no                          & 7.8      & 13.1     & 16.2      \\
                                      & enhanced                    & 7.8      & 12.9     & 15.8     \\
                                      & enhanced+noisy              & 7.8      & 12.9     & 15.6    \\ \midrule
    \multirow{3}{*}{Conformer\_MCT}   & no                          & 6.9   & 12.1      & 13.7  \\
                                      & enhanced                    & 6.7      & 12.0     & 13.5      \\
                                      & enhanced+noisy              & 6.7     & 11.8     & 13.3      \\ \bottomrule
    \end{tabular}
\end{table*}

\begin{table*}[t]
\caption{The impacts of the joint training with and without GAN on SA-ASR pipeline. Results are in CER[\%]}
\label{jointtrainingtable}
    \begin{tabular}{cccccc}
    \toprule
    \multirow{2}{*}{SE}      & \multirow{2}{*}{ASR}         & \multirow{2}{*}{joint training with GANs} & \multicolumn{3}{c}{CER{[}\%{]}} \\ \cline{4-6} 
                             &                              &                                          & clean    & match    & unmatch   \\ \midrule
    \multirow{6}{*}{SA\_SEGAN} & \multirow{2}{*}{baseline}    &     no                                 &    12.8      &  18.7        & 25.3           \\
                             &                            & yes                                      &   \textbf{12.8}       & \textbf{18.7}         & \textbf{24.8}       \\ \cline{2-6} 
                             & \multirow{2}{*}{Transformer} &     no                                 &  \textbf{7.0}    & 12.4          & 15.6           \\
                             &                            & yes                                      &    7.2      & \textbf{12.4}          & \textbf{15.5}           \\ \cline{2-6} 
                             & \multirow{2}{*}{Conformer}   &      no                                  & \textbf{6.8}         & 11.9          & 13.3          \\
                             &                            & yes                                      & 6.9      & \textbf{11.8}     & \textbf{13.0}      \\ \bottomrule
    \end{tabular}
\end{table*}

Secondly, we train SA\_SEGAN with the training data contaminated by ``matched" intrusions in Table \ref{matchunmatch} to enhance the noisy speech. Then the enhanced features are used for the downstream ASR task. Importantly, the ASR models are taken over from the same well-trained model as in Table \ref{table1}, which means that the enhancement front-end and the ASR back-end are trained separately by different objectives. As exhibited in Table \ref{tableenhance}, the enhancement module tremendously improves the performance of the ASR component, which is trained by the clean data merely. Compared to Table \ref{table1}, it outperforms all of the three ASR modules (baseline, Transformer, Conformer) without the enhancement front-end. The improvement achieved in the ``matched" dataset is more remarkable than that achieved on the ``unmatched" test set. For instance, it outperforms the Conformer without the enhancement module by 46.0\% in the ``matched" test set while 9.1\% in the ``unmatched" test set. This difference is due to that the SA\_SEGAN is trained with the ``matched" intrusions and can enhance the data contaminated by the same intrusions better during the test. All these improvements confirm the efficacy of the enhancement module for improving the robustness of the ASR system. Nevertheless, improving the robustness of the framework in unseen noisy environments still remains to be a challenge. Additionally, the speech enhancement module deteriorates the performer of the ASR\_MCT network, which stays in accordance with the observations in \cite{donahue2018exploring}. This degradation may be derived from the latent distortions caused by the overtraining of the enhancement module.

To remedy the deterioration of the performance of the ASR\_MCT, we retrain the network with the enhanced features. Assuming that the network may also benefit from the knowledge of the noisy features, we also experiment with ingesting both enhanced and noisy features. Results are displayed in Table \ref{retraining}. Either the Transformer\_MCT model or the Conformer\_MCT model is initialized from the existing well-trained MCT checkpoints respectively, setting the additional parameters to zero to ensure the fair training start. As presented in Table \ref{retraining}, the retraining with the enhanced features improves the performance in both ``matched" and ``unmatched" cases, and the retraining with both enhanced and noisy features improves the performance slightly further. 

Lastly, we jointly train the whole scheme with and without adversarial training according to Eq. \ref{jointtraining}. In the framework, the enhancement front-end is initialized from the generator (G component) of SA\_SEGAN, the ASR back-end is initialized from the ASR\_MCT checkpoint (without retraining), and the adversarial module is initialized from the discriminator (D component) of SA\_SEGAN. When the adversarial module participates in the training, we set the magnitude of the loss function by $\kappa$=6.0 and $\gamma$=0; by contrast, when it participates in the training, we set $\kappa$=6.0 and $\gamma$=3.0. Results are presented in Table \ref{jointtrainingtable}. Compared to Table \ref{table1}, the joint training mitigates the distortion problem existing in the MCT strategy. Additionally, the participance of the adversarial training improves the performance further; and exceeds the performance of retraining with both enhanced and noisy features in either Transformer or Conformer case. Taking Conformer for example, compared to Conformer trained with clean data merely, the adversarial joint training yields 63.8\% relative and 59.0\% relative improvements on ``matched" and ``unmatched" datasets, respectively; meanwhile, the adversarial joint training outperforms the MCT strategy by 2.5\% relative and 5.2\% relative on ``matched" and ``unmatched" datasets, separately. These results indicate the efficacy of the adversarial joint training in improving the robustness of the end-to-end ASR scheme. 

\begin{figure}[tb]
\centering
\includegraphics[scale=0.42]{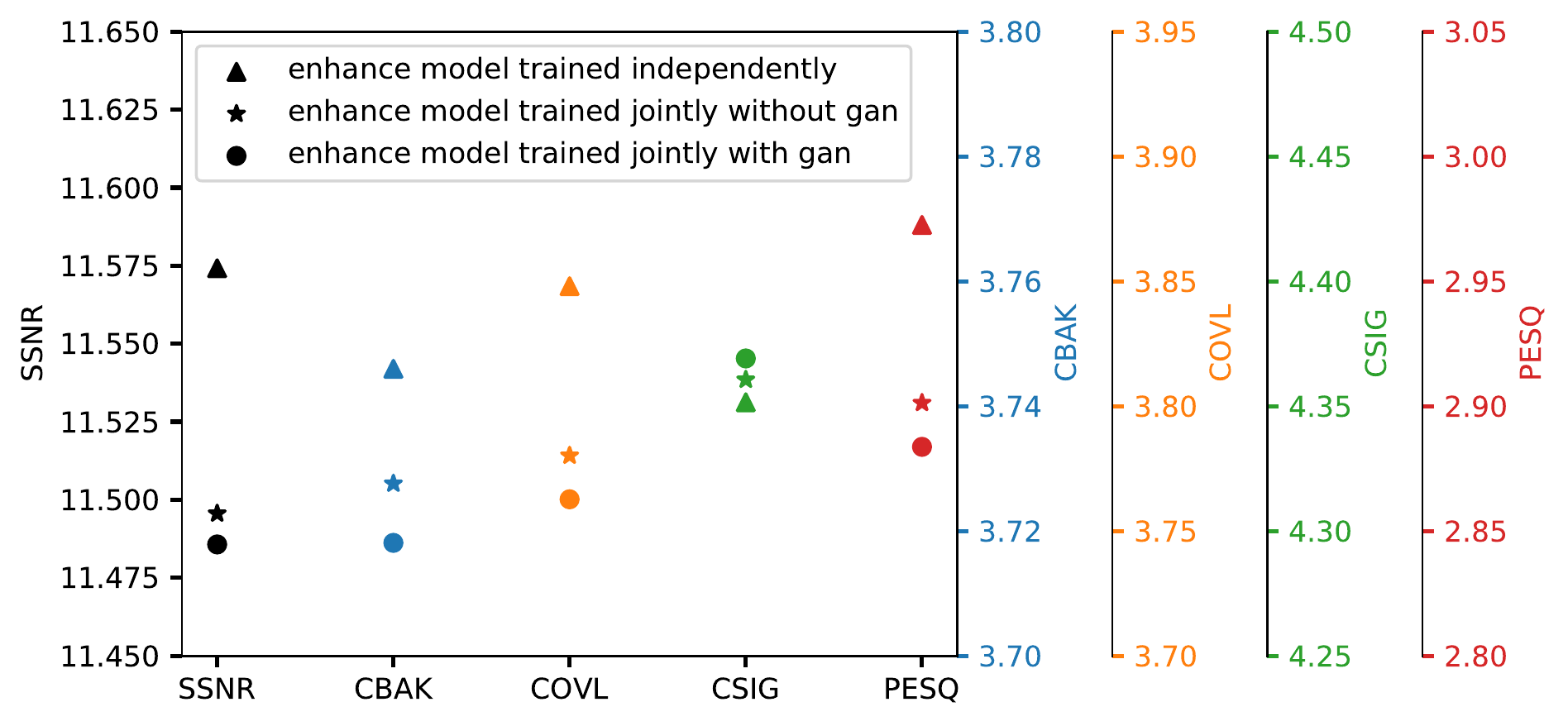}
\caption{Comparisons of different enhancement models' performance under different training conditions.}
\label{evaluation}
\end{figure}

\section{Discussion}
To analyse the difference between these enhancement modules that are trained independently, jointly without GANs, and jointly with GANs, we quantify their performance on the following five objective criteria (the higher the better):
\begin{itemize}
    \item SSNR: Segmental SNR \cite{quackenbush1995objective} (in the range of [0\,,\,$+\infty$))
    \item CBAK: MOS prediction of the intrusiveness of background noises \cite{hu2007evaluation} (in the range of [1\,,\,5])
    \item CSIG: MOS prediction of the signal distortion attending only to the speech signal \cite{hu2007evaluation} (in the range of [1\,,\,5])
    \item COVL: MOS prediction of the overall effect \cite{hu2007evaluation} (in the range of [1\,,\,5])
    \item PESQ: Perceptual evaluation of speech quality, using the wide-band version recommended in ITU-T P.862.2 \cite{rec2005p} (in the range of [-0.5\,,\,4.5])
\end{itemize}

All criteria are computed based on the implementation in \cite{loizou2013speech}, available at the publisher website\footnotemark.
\footnotetext{\href{https://www.crcpress.com/downloads/K14513/K14513_CD_Files.zip}{https://www.crcpress.com/downloads/K14513/\\K14513\_CD\_Files.zip}}
We quantify the performance of the enhancement front-end that is trained independently, trained jointly with and without GANs in case of Transformer scheme. As exhibited in Fig. \ref{evaluation}, the joint training disgrades the enhancement module's performance on SSNR, CBAK, COVL, and PESQ, except for CSIG. It is safe to draw two conclusions from this result. First, these results suggest that these objective criteria cannot indicate the suitability of the enhanced data for ASR task, which verifies the assertion that the independent training leads the enhancement module into the sub-optimum easily. Second, the opposite trend on CSIG proves the assumption that the joint training strategy can mitigate the unseen distortion introduced by the handcrafted loss function. Another phenomenon which is worth nothing is that the discrepancies on CBAK and SSNR reveals the conflicts between erasing the noise contamination and averting the speech distortion. Therefore, the equilibrium between these two goals is critical. The experimental results in Section \ref{results} validate the efficacy of the adversarial joint training with a global discriminant guide for reaching the equilibrium point.

\section{Conclusion}
In this paper, we propose an adversarial joint training framework with the self-attention mechanism to boost the noise robustness of the end-to-end ASR system. The jointly compositional scheme consists of an enhancement front-end, a recognition back-end, and the discriminant network. A highlight of this proposed framework is the discriminant component first acts as the guide of the enhancement front-end training; afterwards, it participates in the adversarial joint training as the global instructor, which leads the enhancement front-end to output appropriate enhanced features for the downstream ASR task. Experimental results validate the efficacy of the proposed adversarial joint training strategy. The next work plan is to investigate different framework architectures and training strategies for further improved performance.

\begin{backmatter}



\section*{Abbreviations}
ASR: automatic speech recognition; SA\_ASR: sela-attention automatic speech recognition; GANs: generative adversarial networks; LSGAN: least-squares generative adversarial networks; SE: speech enhancement; SEGAN: speech enhancement generative adversarial networks; SA\_SEGAN: self-attention speech enhancement generative adversarial networks; CER: character error rate; PReLUs: parametric rectified linear units; FFN: feed-forward network; MHSA: multi-head self-attention module; GLU: gated linear units; STFT: short-time Fourier transform; LSTM: long short-term memory; BLSTM: bidirectional long short-term memory; CTC: connectionist temporal classification; RNN: recurrent neural network; CNN: convolutional neural network; MCT: multi-conditional training.

\section*{Availability of data and materials}
The dataset is the open source Mandarin speech corpus, AISHELL-1 \cite{bu2017aishell} and can be found under the following link: \href{http://www.aishelltech.com/kysjcp}{http://www.aishelltech.com/kysjcp}.


\section*{Competing interests}
The authors declare that the research was conducted in the absence of any commercial or financial relationship that could be construed as potential competing interests.


\section*{Authors' contributions}
Li, L. conceptualised the study. Li, L., Kang, Y., and Shi, Y. executed the experiments. All authors did literature analysis, manuscript preparation, editing, and proofreading, and approved the final manuscript. 



\bibliographystyle{bmc-mathphys} 
\bibliography{bmc_article}      

\end{backmatter}
\end{document}